\documentclass[twocolumn,prl,superscriptaddress,showpacs]{revtex4}
\usepackage{amsmath}

\newcommand{\ket}[1]{\left | \, #1 \right \rangle}
\newcommand{\bra}[1]{\left \langle #1 \, \right |}
\newcommand{\proj}[1]{\ket{#1}\bra{#1}}

\newcommand{\half}{\mbox{$\frac{1}{2}$}}
\newcommand{\third}{\mbox{$\frac{1}{3}$}}
\newcommand{\sixth}{\mbox{$\frac{1}{6}$}}

\begin{document}
\title{Fidelity of Single Qubit Maps}
\author{Mark D. Bowdrey}\email{mark.bowdrey@qubit.org}
\affiliation{Centre for Quantum Computation, Clarendon
Laboratory,University of Oxford, Parks Road, OX1 3PU, United
Kingdom}
\author{Daniel K. L. Oi}\email{daniel.oi@qubit.org}
\affiliation{Centre for Quantum Computation, Clarendon
Laboratory,University of Oxford, Parks Road, OX1 3PU, United
Kingdom}
\author{Anthony J. Short}\email{tony.short@qubit.org}
\affiliation{Centre for Quantum Computation, Clarendon
Laboratory,University of Oxford, Parks Road, OX1 3PU, United
Kingdom}
\author{Jonathan A. Jones}\email{jonathan.jones@qubit.org}
\thanks{to whom correspondence should be addressed at the
Clarendon Laboratory} \affiliation{Centre for Quantum Computation,
Clarendon Laboratory,University of Oxford, Parks Road, OX1 3PU,
United Kingdom} \affiliation{Oxford Centre for Molecular Sciences,
New Chemistry Laboratory,University of Oxford, South Parks Road,
OX1 3QT, United Kingdom}
\date{March 15, 2001}
\pacs{03.67.-a}
\begin{abstract}
We give a simple way of characterising the average fidelity
between a unitary and a general operation on a single qubit which
only involves calculating the fidelities for a few pure input
states.
\end{abstract}
\maketitle

Consider transforming a quantum system by a desired operator, $U$.
In practice, we may not be able to implement $U$ exactly but may
actually apply a map $\mathcal{S}$ instead. In general, the map
$\mathcal{S}$ is a superoperator. To gauge how closely
$\mathcal{S}$ approximates $U$ (or \emph{vice versa}), we require
a measure of how close the output states are to each other,
\emph{i.e.} given identical input states, $\rho_{in}$, how does
$U[\rho_{in}]$ compare with $\mathcal{S}[\rho_{in}]$. For the most
general case where both output states are given by density
operators, we can define their fidelity as \cite{uhlmann76},
\begin{equation}
F(\rho_{1},\rho_{2})=\left(\text{Tr}\left(\sqrt{\sqrt{\rho_{1}}\rho_{2}\sqrt{\rho_{1}}}\,\right)\right)^2,
\end{equation}
which simplifies to
\begin{equation}
F(\proj{\psi},\rho)=\text{Tr}\left(\proj{\psi}\rho\right),
\end{equation}
in the case where at least one of the states is pure
\cite{bruss98}. This situation arises naturally when $U$ is
unitary and the input state, $\rho_{in}$, is pure, as in this case
$U[\rho_{in}]$ will also be pure.

In the case where our system is a single qubit, the allowed
operations are restricted to affine contractions of the Bloch
Ball. We will consider the case where $U$ is a desired unitary
transformation (rotation) of the Bloch ball, and $\mathcal{S}$ is
a superoperator. It is assumed that $\mathcal{S}$ is a linear,
trace-preserving map on the space of single qubit density
operators \cite{ruskai01}. We will define $\bar{F}$ as the average
fidelity over all pure input states,
\begin{equation}\label{eqDefFid}
\bar{F}=\frac{1}{4\pi} \int\! F_{\proj{\psi}} \,\text{d}\Omega,
\end{equation}
where
\begin{equation}
F_{\proj{\psi}}=\text{Tr}\bigl(
U\proj{\psi}U^{\dagger}\mathcal{S}\left[\proj{\psi}\right]\bigr).
\end{equation}
The state $\proj{\psi}$ can be expressed in the basis of the Pauli
spin matrices
\begin{equation}
\begin{split}
\proj{\psi} & = \half\left(\mathbf{1} + \left(
\begin{array}{cc}
\cos\theta & \sin\theta e^{-i\phi} \\
 \sin\theta e^{i\phi} & -\cos\theta
\end{array}
\right)
\right)
\\
&= \frac{\sigma_{0}}{2}+\sin\theta\,\cos\phi\,\frac{\sigma_{x}}{2}
+\sin\theta\,\sin\phi\,\frac{\sigma_{y}}{2}+\cos\theta\,\frac{\sigma_{z}}{2}\\
&= \!\!\!\!\!\sum_{j=0,x,y,z}\!\!\!\!\! c_{j}(\theta,\phi)
\frac{\sigma_{j}}{2}.
\end{split}
\end{equation}
Hence Eq.~(\ref{eqDefFid}) can be expressed as
\begin{widetext}
\begin{equation}
\begin{split}
\bar{F} & = \frac{1}{4\pi}\int_{\theta=0}^{\pi}
\int_{\phi=0}^{2\pi} \text{Tr}\left( U\sum_{j}c_{j}(\theta,\phi)
\frac{\sigma_{j}}{2}U^{\dagger}\,
\mathcal{S}\!\left[\sum_{k}c_{k}(\theta,\phi) \frac{\sigma_{k}}{2}\right]\right)
\sin{\theta}\, \text{d}\phi\, \text{d}\theta\\
        & =
\sum_{jk} \left(
  \frac{1}{4\pi}\int_{\theta} \int_{\phi} c_{j}c_{k}\sin\theta\, \text{d}\theta\, \text{d}\phi
\right) \text{Tr}\left( U\frac{\sigma_{j}}{2} U^{\dagger} \,
\mathcal{S} \!\left[ \frac{\sigma_{k}}{2} \right] \right)
\end{split}
\end{equation}
\end{widetext}
where we have used the linearity of $U$ and $\mathcal{S}$. The
integrals of the coefficients, $c_{jk}$, are easily evaluated by
symmetry: the cross terms vanish~\cite{bowdrey01} leaving the
simpler expression
\begin{equation}
\begin{split}
\bar{F} & =
\sum_{jk}\left(\frac{2\delta_{j0}\delta_{k0}+\delta_{jk}}{3}\right)
\text{Tr}\left( U\frac{\sigma_{j}}{2} U^{\dagger} \, \mathcal{S} \!\left[ \frac{\sigma_{k}}{2} \right] \right)\\
        & =
\text{Tr}\left(U\frac{\sigma_{0}}{2}U^{\dagger}\,
\mathcal{S}\!\left[\frac{\sigma_{0}}{2}\right]\right)+
\third\!\!\!\!\sum_{j=x,y,z}\!\!\!\!\text{Tr}\left(U\frac{\sigma_{j}}{2}U^{\dagger}\, \mathcal{S}\!\left[\frac{\sigma_{j}}{2}\right]\right)\\
        & =
\half+
\third\!\!\!\!\sum_{j=x,y,z}\!\!\!\!\text{Tr}\left(U\frac{\sigma_{j}}{2}U^{\dagger}\,
\mathcal{S}\!\left[\frac{\sigma_{j}}{2}\right]\right)
\end{split}
\end{equation}
where we have used the fact that $\mathcal{S}[\sigma_0/2]$ is a
density matrix and thus has unit trace.  In order to express the
average fidelity in terms of states, and to give a more intuitive
picture of the above expression, we use the substitutions,
\begin{equation}
\begin{split}
\frac{\sigma_{j}}{2} & =
\frac{\sigma_0+\sigma_{j}}{2}-\frac{\sigma_0}{2}
                       = \rho_{j}-\rho_{0}\\
                     & = \frac{\sigma_0}{2} - \frac{\sigma_0-\sigma_{j}}{2}
                       = \rho_{0} - \rho_{-j},
\end{split}
\end{equation}
where $\rho_{\pm j}$ represents a pure state in the $\pm
j$-direction and $\rho_{0}$ is the maximally mixed state. This
gives the two equivalent expressions,
\begin{equation}\label{eqFidAv}
\bar{F} = \half+\third\!\!\!\!\sum_{j=x,y,z}\!\!\!\!
\text{Tr}\left(U\rho_{j}U^{\dagger}\,
\mathcal{S}\!\left[\rho_{j}\right]- U\rho_{j}U^{\dagger}\,
\mathcal{S}\!\left[\rho_{0}\right]\right)
\end{equation}
\begin{equation}\label{eqFidAv2}
\bar{F} = \half+\third\!\!\!\!\sum_{j=x,y,z}\!\!\!\!
\text{Tr}\left(U\rho_{-j}U^{\dagger}\,
\mathcal{S}\!\left[\rho_{-j}\right]- U\rho_{-j}U^{\dagger}\,
\mathcal{S}\!\left[\rho_{0}\right]\right)
\end{equation}
and taking their average yields,
\begin{widetext}\begin{equation}\begin{split}
\bar{F} & = \half+\sixth\!\!\!\!\sum_{j=x,y,z}\!\!\!\!
\text{Tr}\left(U\rho_{j}U^{\dagger}\,
\mathcal{S}\!\left[\rho_{j}\right]+ U\rho_{-j}U^{\dagger}\,
\mathcal{S}\!\left[\rho_{-j}\right]-
U(\rho_{j}+\rho_{-j})U^{\dagger}\,
\mathcal{S}\!\left[\rho_{0}\right]\right)\\
        & =
\half+\sixth\!\!\!\!\sum_{j=x,y,z}\!\!\!\!
\text{Tr}\left(U\rho_{j}U^{\dagger}\,
\mathcal{S}\!\left[\rho_{j}\right]+ U\rho_{-j}U^{\dagger}\,
\mathcal{S}\!\left[\rho_{-j}\right]- 2\, U\rho_{0}U^{\dagger}\,
\mathcal{S}\!\left[\rho_{0}\right]\right)\\
        & =
\half+\sixth\!\!\!\!\sum_{j=x,y,z}\!\!\!\!\left(
\text{Tr}\left(U\rho_{j}U^{\dagger}\,
\mathcal{S}\!\left[\rho_{j}\right]\right)+
\text{Tr}\left(U\rho_{-j}U^{\dagger}\,
\mathcal{S}\!\left[\rho_{-j}\right]\right)-1
\right)\\
        & =
\sixth\!\!\!\!\sum_{j=\pm x,\pm y,\pm z}\!\!\!\!\left(
\text{Tr}\left(U\rho_{j}U^{\dagger}\,
\mathcal{S}\!\left[\rho_{j}\right]\right) \right).
\end{split}\end{equation}
\end{widetext}
Hence, the fidelity of the superoperator $\mathcal{S}$ with the
unitary operator $U$ can be calculated by simply averaging the
fidelities of the six axial pure states on the Bloch sphere,
$\{\rho_{+x}$, $\rho_{-x}$, $\rho_{+y}$, $\rho_{-y}$, $\rho_{+z}$,
$\rho_{-z}\}$. In the case where $\mathcal{S}$ is unital
($\mathcal{S}[\rho_0]=\rho_0$) it can be seen from
Eq.~\ref{eqFidAv} that this reduces to an average over only three
states, $\{\rho_{+x}$, $\rho_{+y}$, $\rho_{+z}\}$; similarly
Eq.~\ref{eqFidAv2} shows that the average can be taken over
$\{\rho_{-x}$, $\rho_{-y}$, $\rho_{-z}\}$.

\begin{acknowledgments}We thank E.~Galv\~{a}o and L.~Hardy for
helpful conversations. M.D.B. and A.J.S. thank EPSRC (UK) for a
research fellowship.  D.K.L.O thanks CESG (UK) for financial
support. J.A.J. is a Royal Society University Research Fellow.
This work is in part a contribution from the Oxford Centre for
Molecular Sciences, which is supported by the UK EPSRC, BBSRC, and
MRC.
\end{acknowledgments}


\begin{thebibliography}{9}

\bibitem{uhlmann76}
A. Uhlmann, Rep. Math. Phys. \textbf{9}, 273 (1976).
\bibitem{bruss98}
D. Bruss, D. P. DiVincenzo, A. Ekert, C. A. Fuchs, C. Macchiavello
and J. A. Smolin, Phys. Rev. A. \textbf{57}, 2368 (1998).
\bibitem{ruskai01}
M. B. Ruskai, S. Szarek, and E. Werner, LANL e-print
quant-ph/0101003.
\bibitem{bowdrey01}
M. D. Bowdrey and J. A. Jones, LANL e-print quant-ph/0103060.

\end{thebibliography}
\end{document}